\documentclass[prd,twocolumn,showpacs,preprintnumbers]{revtex4} 

\input{epsf}
\def\to{\rightarrow}

\def\AJ{{\it Ap. J.} }

\def\ASAS{{\it Astron. and Astrophys.} }

\def\IJMP{{\it Int. J. Mod. Phys.} }

\def\PL{{\it Phys. Lett.} }
\def\PR{{\it Phys. Rev.} }
\def\PRL{{\it Phys. Rev. Lett.} }

\def\frac#1#2{{\textstyle{{#1}\over {#2}}}}

\def\lsim{\mathrel{\rlap{\lower4pt\hbox{\hskip1pt$\sim$}}
    \raise1pt\hbox{$<$}}}
\def\gsim{\mathrel{\rlap{\lower4pt\hbox{\hskip1pt$\sim$}}
    \raise1pt\hbox{$>$}}}
\def\sqr#1#2{{\vcenter{\vbox{\hrule height.#2pt
         \hbox{\vrule width.#2pt height#1pt \kern#1pt
         \vrule width.#2pt}
         \hrule height.#2pt}}}}

\def\beq{\begin{equation}}
\def\eeq{\end{equation}}
\def\beqa{\begin{eqnarray}}
\def\eeqa{\end{eqnarray}}

\begin{document}

\title{Varying Electromagnetic Coupling and Primordial Magnetic Fields}

\author{O. Bertolami and R. Monteiro}

\vskip 0.2cm

\affiliation{Departamento de F\'\i sica, Instituto Superior T\'ecnico \\
Av. Rovisco Pais 1, 1049-001 Lisboa, Portugal}

\vskip 0.2cm

\affiliation{E-mail addresses: orfeu@cosmos.ist.utl.pt; ricardo58monteiro@gmail.com}

\vskip 0.5cm

\date{\today}

\begin{abstract}
We study the effect of variations of the electromagnetic coupling on the process of generation of primordial magnetic fields. We find that only through a significant growth of the electromagnetic coupling minimum seed fields can be produced. We also show that, if through some process in the early Universe the photon acquires a mass that leads, thanks to inflation, to the generation of primordial magnetic fields, then the influence of variations of the electromagnetic coupling amounts essentially to the results due to the photon effective mass alone.
 
\vskip 0.5cm
 
\end{abstract}

\pacs{98.80.Cq, 98.65.Es \hspace{2cm}Preprint DF/IST-1-2005}

\maketitle
\section{Introduction}

The existence of feeble magnetic fields extending over large distances is now firmly established. In intra-galactic mediums, as in our Milky Way, fields with strength $\sim \mu$G can be coherent over distances similar to the dimensions of the galaxies themselves. Recent observations indicate also the existence of similar magnetic fields on much larger scales, as in clusters or even superclusters, with strengths in the range of $\mu$G (see eg. Ref. \cite{Gi} for a review).

Despite the great interest around this subject for at least the last 20 years, the most important question remains essentially open: what is the origin of these fields?

Another surprising fact about this fields is that their energy density, $\rho_B=B^2/8\pi$, is of the same order of magnitude of the Cosmic Microwave Background Radiation (CMBR) energy density, namely $\rho_\gamma=1.86\times 10^{-33}h^2$~g$\,$cm$^{-3}$, where $h$ parametrizes the uncertainty on the Hubble constant $H_0=100\,h$~km$\,$s$^{-1}\,$Mpc$^{-1}$ and from the observations $h\simeq 0.7$. Since the Universe behaved as good conductor, very early on in its history \cite{TW}, the magnetic flux is frozen, so that the ratio $r=\rho_B/\rho_\gamma$ is almost constant, which provides a convenient measure of the magnetic field strength. For galaxies, $r \sim 1$.

The dynamo mechanism, whose several adaptations are also discussed in Ref. \cite{Gi}, is capable of amplifying galactic magnetic fields, due to the differential rotation of turbulent plasma in the galaxy, by an exponential factor. However, it is clear that seed magnetic fields are required. Its minimal magnitude is estimated to be $r\sim 10^{-34}$ in order to reach the present field strength. On the other hand, the observed magnetic fields may have its origin in the compression of the magnetic flux that may be trapped in the gas during the collapse of the protogalactic cloud. This process requires seed fields of order $r\sim 10^{-8}$.

The fact that magnetic fields seem to exist over quite large scales has led to the idea that inflation could be the mechanism behind its origin. Inflation allows the emergence of fields coherent over large distances from quantum fluctuations and prevents dissipation effects, due to the absence of charged particles. During inflation, conductivity is very poor. However, the conformal invariance of the $U(1)$ gauge theory for electromagnetism prevents the gravitational field from producing photons, as if conductivity were high. Under these conditions inflation significantly suppresses field fluctuations yielding $r\sim 10^{-104} \lambda_{\mathrm{Mpc}}^{-4}$, where $\lambda_{\mathrm{Mpc}}\equiv\lambda /\mathrm{Mpc}$. 

In Ref. \cite{TW}, it is proposed the breaking of the conformal invariance of electromagnetism via the coupling to gravity in different ways. In spite of the drawback represented by the breaking of the gauge symmetry, the addition of a mass-like term $RA_\mu A^\mu$ to the free Lagrangian of electromagnetism yields interesting results. The origin of an effective mass for the photon during inflation has been speculated since then. In Ref. \cite{BM}, it is suggested that the photon mass is acquired from the spontaneous breaking of the Lorentz symmetry in the context of field theories arising from string field theory (see \cite{KB} and references therein). In Ref. \cite{AM}, on the other hand, the photon is endowed with a mass by introducing a minimal fundamental length in a transplanckian physics scenario.

An apparently unrelated issue is the indication of variations of the electromagnetic coupling in the late history of the Universe resulting from recent observations of the spectra of quasars \cite{We} (see also Refs. \cite{Ch}). This possibility is, at least in principle, fairly natural in the context of Kaluza-Klein theories \cite{KK} and string theories \cite{ST}. The study of varying coupling cosmologies is usually done by coupling fields to electromagnetism, as proposed by Bekenstein \cite{Be}. Various proposals consider a scalar field \cite{sf}, quintessence \cite{q1}, coupled quintessence fields \cite{q2}, etc. The possible relation between the coupling variation and the production of seed magnetic fields has been proposed through different models, for instance, in the context of electromagnetism coupled with a dilaton \cite{de}.

In this work, we consider a scalar field coupled to electromagnetism, as suggested in Ref. \cite{Be}, and study its influence in the inflationary generation of the magnetic fluctuations. We shall perform this by analyzing the generalized wave equation for the magnetic flux and the equation of motion for the scalar field. We shall study the influence of the coupling evolution together with a photon mass model on the generation of primordial magnetic fields of Ref.~\cite{BM}.


\section{The Evolution of the Magnetic Flux and the Electromagnetic Coupling}

We consider spatially flat Friedmann-Robertson-Walker cosmologies, where the metric in the conformal time $\eta$ is given by:
\begin{equation}
\label{metric}
g_{\mu \nu} = a(\eta)^2  \mathrm{diag}(-1,1,1,1)\;,
\end{equation}
where $a(\eta)$ is the scale factor.

We shall study the generation of seed magnetic fields in a model where the photon has an effective mass and where a homogeneous scalar field $\phi$ is inversely proportional to the square of the electromagnetic coupling. The appropriate effective Lagrangian density, in natural units, is
\begin{equation}
{\mathcal L} = {\mathcal L}_{em} + {\mathcal L}_\phi\;,
\end{equation}
where
\begin{equation}
{\mathcal L}_{em}=- {1 \over 4} \,\phi\, F_{\mu \nu} F^{\mu \nu} + {1 \over 2} M^2 A_\mu A^\mu
\end{equation}
accounts for the electromagnetic field coupled to the scalar field and
\begin{equation}
{\mathcal L}_\phi=- {1 \over 2}\, \partial_\mu \phi \, \partial^\mu \phi - V(\phi)
\end{equation}
is the usual Lagrangian density for the scalar field.

The equation of motion for the scalar field is then
\begin{equation}
\label{eq:varphiF}
\ddot{\phi} + 2\, {\dot{a} \over a} \, \dot{\phi} + a^2\, V'(\phi)+ {1 \over 4}\, a^2 \,F_{\mu \nu} F^{\mu \nu} =0\;,
\end{equation}
where the dot denotes differentiation with respect to the conformal time.

The electromagnetic field obeys the equation of motion
\begin{equation}
\label{eq:eqmovF}
(\partial_\mu \phi) F^{\mu \nu}+\phi \,\nabla_\mu F^{\mu \nu}+ M^2 A^\nu=0
\end{equation}
and the Bianchi identity
\begin{equation}
\partial_\gamma F_{\alpha \beta}+\partial_\beta F_{\gamma \alpha}+\partial_\alpha F_{\beta \gamma}=0 \;.
\end{equation}
The mass term in Eq. (\ref{eq:eqmovF}) may have its origin on the spontaneous breaking of Lorentz invariance from trilinear terms like $TA_\mu A^\mu$, where $T$ is a generic tensor field, arising, for instance, from open string field theory \cite{BM}.
The field strength tensor $F_{\mu \nu}$ is given by
\begin{eqnarray}
F_{\mu \nu}= a(\eta)^2 \left( \begin{array}{cccc} 0&-E_x&-E_y&-E_z \\ E_x&0&B_z&-B_y \\ E_y&-B_z&0&B_x \\ E_z&B_y&-B_x&0 \end{array}\right)\;,
\end{eqnarray}
so that the relevant Maxwell equations read
\begin{eqnarray}
\label{eq:rotE}
{1 \over a^2}\,{\partial \over \partial \eta}(a^2 \vec{B})+\vec{\nabla} \times \vec{E}=0 \;,\\
\label{eq:rotB}
{1 \over a^2}\,{\partial \over \partial \eta}(\phi a^2 \vec{E})-\phi(\vec{\nabla} \times \vec{B})+ M^2\vec{A} =0 \;.
\end{eqnarray}

Taking the curl of Eq. (\ref{eq:rotB}) and using Eq. (\ref{eq:rotE}), we obtain the appropriate wave equation for the magnetic field:
\begin{equation}
\label{eq:evolB}
{1 \over a^2}\,\dot{\phi} \,{\partial \over \partial \eta}(a^2 \vec{B})+{\phi \over a^2}{\partial^2 \over \partial \eta^2}(a^2 \vec{B})-\phi \nabla^2 \vec{B}+ {n \over \eta^2}\vec{B}=0\;,
\end{equation}
where we have introduced, for later convenience, $n \equiv -\eta^2 M^2 a^2$.

Defining the magnetic flux associated with the comoving scale $\lambda= 2 \pi k^{-1}$ as a Fourier transform
\begin{equation}
\vec{F}_k(\eta)\equiv a^2 \int d^3x e^{i\vec{k} \cdot \vec{x}} \vec{B}(\vec{x},\eta)\;,
\end{equation}
the equation for the evolution of the magnetic flux reads:
\begin{equation}
\label{eq:evolFk}
\dot{\phi}\dot{\vec{F}}_k+\phi \ddot{\vec{F}}_k+k^2 \vec{F}_k + {n \over \eta^2}\vec{F}_k=0\;.
\end{equation}
If one admits that the magnetic flux does not change its direction appreciably during the relevant period of time, the result for modes well outside the horizon ($a\lambda\gg H^{-1}$ or $|k\eta|\ll 1$) is
\begin{equation}
\label{eq:evolF}
\dot{\phi}|\dot{\vec{F}}_k|+ \phi |\ddot{\vec{F}}_k|+{n \over \eta^2}|\vec{F}_k|=0\;.
\end{equation}
The magnetic flux is a measure of the energy density associated with the comoving scale. If $\rho_B(k)=k\,d\rho_B/dk$, one has $\rho_B(k)\propto |\vec{F}_k|^2/a^4$. It is clear from Eq. (\ref{eq:evolF}) that the effect of the variation of the electromagnetic coupling is similar to a dissipation term.

It is helpful in the following to discuss the most salient aspects of the photon mass model of Refs. \cite{TW,BM}, which assumes a constant coupling. 

The parameter $n$ is negative and constant during each period of the evolution of the Universe. In Ref. \cite{TW}, $M^2$ equals $R=6\frac{\ddot{a}}{a^3}$ apart from a constant factor, so that $n$ is constant because $a(\eta)$ varies as a power of $\eta$ in each period. In Ref. \cite{BM}, where $M^2=M_L^2a^{-2l}$ and $M_L$ is characteristic of the period, the parameter $l$ associated with the violation of the Lorentz symmetry is chosen such that $n$ does not evolve. As the coupling is assumed constant, setting $\phi=1$, the solution of Eq. (\ref{eq:evolF}) is given by
\begin{equation}
\label{eq:solF}
|\vec{F}_k|\propto \eta^{m_\pm} \quad \mathrm{with} \quad m_\pm={1 \over 2}(1\pm \sqrt{1-4n})\; .
\end{equation}

During the de Sitter (dS) phase, the scale factor varies as $a(\eta)\propto -\frac{1}{\eta}$, so that the non-dissipative power is $p={m_-}_{dS}\le 0$, and the energy density of the Universe corresponds to $\rho_{TOT}=M_{GUT}^4$, where $M_{GUT}$ is an unification scale. In the Reheating (RH) phase, $a(\eta)\propto \eta^2$ and the convenient power is $q={m_+}_{RH}\ge 1$. The energy density varies then approximately as $\rho_{TOT}\simeq T^8 T_{RH}^{-4} \propto a^{-3}$, where $T_{RH}$ is the reheating temperature. However, in Ref. \cite{TW} it is assumed that the early Universe is a good conductor starting at a period characterized by the temperature $T_\ast$. Afterwards, the magnetic flux is frozen so that $\rho_B(k)\propto a^{-4}$. In the Radiation Dominated and Matter Dominated eras, one also has $\rho_{TOT}\sim \rho_\gamma \propto a^{-4}$ and so the ratio $r=\rho_B(k) / \rho_{TOT}$ remains constant. Accordingly, we may write
\begin{eqnarray}
\label{eq:rinicial}
\lefteqn{r=\left( {\rho_B(k) \over \rho_{TOT}} \right)_{a=a_1} e^{-2\,N(\lambda)\,(p+2)} \times {}}\nonumber \\ && {}\times \left({M_{GUT} T_{RH} \over {T_\ast}^2}\right)^{\frac{4}{3}(q-1)} \left({T_\ast \over T_{RH}}\right)^{-\frac{8}{3}},
\end{eqnarray}
where $a_1$ denotes the scale factor in the instant of the crossing of the horizon by the fluctuation. For the first factor one has
\begin{equation}
\left( {\rho_B(k) \over \rho_{TOT}} \right)_{a=a_1}\simeq \left( {M_{GUT} \over M_P} \right)^4 \;,
\end{equation}
where $M_P$ is the Planck mass and $N(\lambda)$ is the number of \emph{e-foldings} between the horizon crossing and the end of the dS period, given by \cite{EU}:
{\setlength\arraycolsep{2pt}
\begin{eqnarray}
\label{eq:Nlambda}
N(\lambda)&=&45+ {2 \over 3}\ln \left( {M_{GUT} \over 10^{14}\mathrm{~GeV}} \right) + {1 \over 3}\ln \left( {T_{RH} \over 10^{10}\mathrm{~GeV}} \right) +{}\nonumber \\ && {}+\ln\lambda_{\mathrm{Mpc}}\;.
\end{eqnarray}}
Hence, one obtains an explicit formula for $r$ \cite{TW}:
\begin{eqnarray}
\label{eq:r}
\lefteqn{r\simeq (9 \times 10^{25})^{-2(p+2)} \left({M_{GUT} \over M_P}\right)^{\frac{4}{3}(q-p)}\times {}}\nonumber \\ && {}\times \left({T_{RH} \over M_P}\right)^{\frac{2}{3}(2q-p)} \left({T_\ast \over M_P}\right)^{-\frac{8}{3}q} {\lambda_{\mathrm{Mpc}}}^{-2(p+2)}\;.
\end{eqnarray}

Notice that if the Universe had always behaved as a good conductor, that is $p=q=0$, and then $r\sim 10^{-104} \lambda_{\mathrm{Mpc}}^{-4}$. The parameter $n$ allows for $p<0$ and $q>1$, implying that fluctuations are not significantly suppressed.

As the dependence of $r$ on the scale $\lambda_{\mathrm{Mpc}}$ is controlled essentially by the parameter $p$, requiring the magnetic fields to be coherent over different scales allows for a prediction of its value. As referred to in Ref. \cite{Gi}, the typical amplitudes of the magnetic fields in superclusters ($\sim 100$~Mpc) can be one order of magnitude smaller than the few $\mu$G fields observed in galaxies ($\sim 100$~kpc). This would mean that the ratio $r$ is poorly dependent on the coherence scale, leading to $p \simeq -2$.

After this brief discussion of the photon mass model, we shall consider two approaches in the next section: 1) We shall assume a variation for $|\vec{F}_k|$ and look for the corresponding variation of $\phi$; 2) We shall assume a variation for $\phi$ and look for the result in $|\vec{F}_k|$. Preparing for this last approach, in order to have an Ansatz for the evolution of the coupling, we consider the scalar field dynamics.

We neglect in Eq. (\ref{eq:varphiF}) the influence of the electromagnetic term. Assuming that $V(\phi)=0$, the solution for the dS period, such that $a\propto e^{Ht}\propto -\frac{1}{\eta}$, is given by
\begin{equation}
\label{eq:phic3}
\phi(\eta)=\bar{\phi}+C \eta^3\;,
\end{equation}
where $\bar{\phi}$ and $C$ are integration constants. However, the potential of a scalar field during the exponential inflation should include a term due to the Hawking temperature $T_H=H/2\pi$ \cite{RB,BR}, that is
\begin{equation}
V(\phi)=H^2 \phi^2 \;,
\end{equation}
which leads to the solution
\begin{equation}
\phi(\eta)=C_1 \eta + C_2 \eta^2\;,
\end{equation}
where $C_1$ and $C_2$ are integration constants. Notice that a rescaling of the ground state of the scalar field yields a new constant term in the solution
\begin{equation}
\label{eq:phic1c2}
\phi(\eta)=\bar{\phi}+C_1 \eta + C_2 \eta^2\;.
\end{equation}


\section{Solutions}

We start the analysis of Eq. (\ref{eq:evolF}) supposing that the photon mass is zero, so that $n=0$. This means we focus now on the influence of the variation of the electromagnetic coupling alone.

Amongst the considerations that led to Eq. (\ref{eq:r}), it has been assumed that $\phi$ was constant, however that expression is still valid whether the magnetic flux $|\vec{F}_k|$ evolves as a power of $\eta$ during the dS and the RH periods. Let us now analyze the corresponding evolution of $\phi$ in case an amplification can be obtained. As $n=0$ here, the evolution of $\phi$ is parametrized by $p$ e $q$. Eq. (\ref{eq:evolF}) then reads
\begin{equation}
\label{eq:phiFconst}
\phi |\dot{\vec{F}}_k| = \mathrm{constant}\;.
\end{equation}
Excluding the trivial solution of constant $|\vec{F}_k|$, we admit that, as before, it varies as a power of $\eta$ and hence
\begin{equation}
\label{eq:evolFeta}
\phi |\vec{F}_k| \propto \eta\;.
\end{equation}
Since the relation between the conformal time and the scale factor is different for the periods dS and RH, we must analyze them separately.

In the dS phase, Eq. (\ref{eq:evolFeta}) in terms of the scale factor reads $\phi |\vec{F}_k| \propto a^{-1}$. Using the previous notation, we consider that $|\vec{F}_k| \propto a^{-p}$, so that the appropriate variation of $\phi$ is
\begin{equation}
\label{eq:phisoldS}
\phi \propto a^{p-1}\;.
\end{equation}
A physical solution should be such that $p<0$ meaning that a greater dissipation of $\phi$ is expected during the exponential inflationary period.

For the RH phase, we have $\phi |\vec{F}_k| \propto a^\frac{1}{2}$ as long as $T>T_\ast$. Considering $|\vec{F}_k| \propto a^\frac{q}{2}$, it then follows that
\begin{equation}
\label{eq:phisolRH}
\phi \propto a^{\frac{1}{2}(1-q)}\;.
\end{equation}
It is now possible to have a growing flux $|\vec{F}_k|$ for $q=1$, while keeping $\phi$ constant. This is similar to the solution Eq. (\ref{eq:solF}) for $m_+=1$. If however $q>1$, attenuation of $\phi$ follows.

Combining the two periods, the variation of $\phi$ that is compatible with Eq. (\ref{eq:r}) is given by
\begin{equation}
\phi = \phi_0 \left({a_{dS} \over a_0} \right)^{p-1} \left({a_{\phi} \over a_{dS}} \right)^{\frac{1}{2}(1-q)}\;,
\end{equation}
where $a_{\phi}$ is the scale factor in the moment $\phi$ becomes constant. For simplicity, let us assume that $a_{\phi}=a_\ast$. Furthermore, it is necessary to introduce a $a_0$, instead of $a_1$, as the variation of $\phi$ cannot depend on the wavelength of the various modes. Since the modes with greater wavelength are the first ones to cross the horizon, the scale to which $a_0$ should be related with is the largest coherence scale of the observed magnetic fields. As these fields are not negligible at the scale of superclusters ($\sim 100$~Mpc), we take for reference the scale of the present observable Universe (Gpc). Notice that, if $p=q=0$, we do not recover $\phi=\mathrm{constant}$ from the expression above, as that case was excluded when assuming that the right hand side of Eq. (\ref{eq:phiFconst}) was non-vanishing.

In a similar way to the steps from Eq. (\ref{eq:rinicial}) to Eq. (\ref{eq:r}), we obtain
\begin{eqnarray}
\label{eq:varphi}
\lefteqn{\phi\simeq\phi_0 \,(9 \times 10^{28})^{p-1} \left({M_{GUT} \over M_P}\right)^{\frac{2}{3}(p-q)}\times {}}\nonumber \\ && {}\times \left({T_{RH} \over M_P}\right)^{\frac{1}{3}(p-2q+1)} \left({T_\ast \over M_P}\right)^{\frac{4}{3}(q-1)}\;.
\end{eqnarray}
The growth factor of the square of the electromagnetic coupling is then given by $(\phi/\phi_0)^{-1}$.

In Table~\ref{tab:rn0phi}, we find some numerical examples of the effectiveness of the photon mass mechanism on the amplification of the fluctuations (columns with $n_{\bar{\phi}}$ and $\bar{\phi}$ should be ignored for now).
\begin{table*}
\begin{ruledtabular}
\begin{tabular}{ccccc|cc|cc|c}
$p$ & $q$ & $M_{GUT}$~(GeV) & $T_\ast$~(GeV) & $T_{RH}$~(GeV) & $n_{\bar{\phi}_{dS}}$ & $n_{\bar{\phi}_{RH}}$ & $\bar{\phi}_{dS}$ & $\bar{\phi}_{RH}$ & $\log_{10} r\,|_{1\mathrm{~Mpc}}$ \\
\hline\hline
 -1 & 2 & $10^{17}$ & $10^{12.3}$ & $10^9$    & -2 & -2 & 1 & 1 & -57 \\
 -1 & 2 & $10^{17}$ & $10^{17}  $ & $10^{17}$ & -2 & -2 & 1 & 1 & -56 \\
 -2 & 3 & $10^{17}$ & $10^{12.3}$ & $10^9$    & -6 & -6 & 1/3 & 1/3 & -13 \\
 -2 & 3 & $10^{17}$ & $10^{17}  $ & $10^{17}$ & -6 & -6 & 1/3 & 1/3 & -8  \\
\hline
 -1.67 & 1 & $10^{17}$ & $10^{13}$ & $10^9$    & -4.46 & 0 & 0.45 &  & -33 \\
 -1.67 & 1 & $10^{17}$ & $10^{17}$ & $10^{17}$ & -4.46 & 0 & 0.45 &  & -24 \\
 -2.08 & 1 & $10^{17}$ & $10^{13}$ & $10^9$    & -6.41 & 0 & 0.31 &  & -16 \\
 -2.08 & 1 & $10^{17}$ & $10^{17}$ & $10^{17}$ & -6.41 & 0 & 0.31 &  & -5  \\
\hline
 -1.67 & 1 & $10^{16}$ & $10^{12.3}$ & $10^9$    & -4.46 & 0 & 0.45 &  & -35 \\
 -1.67 & 1 & $10^{16}$ & $10^{16}  $ & $10^{16}$ & -4.46 & 0 & 0.45 &  & -27 \\
 -2.08 & 1 & $10^{16}$ & $10^{12.3}$ & $10^9$    & -6.41 & 0 & 0.31 &  & -18 \\
 -2.08 & 1 & $10^{16}$ & $10^{16}  $ & $10^{16}$ & -6.41 & 0 & 0.31 &  & -9  \\
\hline
 -1.67 & 1 & $10^{15}$ & $10^{12}$ & $10^9$    & -4.46 & 0 & 0.45 &  & -37 \\
 -1.67 & 1 & $10^{15}$ & $10^{15}$ & $10^{15}$ & -4.46 & 0 & 0.45 &  & -31 \\
 -2.08 & 1 & $10^{15}$ & $10^{12}$ & $10^9$    & -6.41 & 0 & 0.31 &  & -21 \\
 -2.08 & 1 & $10^{15}$ & $10^{15}$ & $10^{15}$ & -6.41 & 0 & 0.31 &  & -13 \\
\end{tabular}
\caption{\label{tab:rn0phi} Values of the ratio $r$ at 1~Mpc scale. We present the values of $n_{\bar{\phi}}$ associated to the parameters $p$ and $q$ according to the photon mass model and the corresponding $\bar{\phi}$. It is assumed that for the different values of $p$ and $q$ across the table a fixed photon mass is taken such that $n_{dS}=n_{RH}=-2$ for reference. Of course, $\bar{\phi}$ is meaningless if $n_{\bar{\phi}}=0$. One should note the strong dependence of the ratio $r$ on $\bar{\phi}$. The choices of the various parameters are the ones from Refs. \cite{TW,BM}.}
\end{ruledtabular}
\end{table*}
The corresponding results from the variation of $\phi$ alone, in the parameters range considered, leads to
\begin{equation}
{(\phi/\phi_0)^{-1} \over r\,|_{1\mathrm{~Mpc}}} \sim 10^{88} \; \mathrm{to} \; 10^{110}\;.
\end{equation}
Hence, the conclusion is that the generation of magnetic fields from inflationary fluctuations based entirely on the evolution of the electromagnetic coupling requires, as expected, an unreasonable change of the electromagnetic coupling.


We have considered a variation for $|\vec{F}_k|$ as a power of the conformal time. Let us now discuss the case with a simple analytical solution in which $\phi$ varies linearly with the conformal time, as allowed by Eq. (\ref{eq:phic1c2}). The solution of Eq. (\ref{eq:phiFconst}) is then
\begin{equation}
\label{eq:solFlnphi}
|\vec{F}_k| = C'_1 + C'_2 \ln \phi\;,
\end{equation}
where $C'_1$ and $C'_2$ are integration constants. This solution is less interesting as the logarithmic variation is much weaker. Nevertheless, it is worthwhile to mention that $|\vec{F}_k|$ may vary by several orders of magnitude if $\phi$ values are close to unity. Assuming $C'_1=0$, then,
\begin{equation}
r = r_0 \left({\ln \phi \over \ln \phi_0}\right)^2= r_0 \left({\ln (1+\varepsilon) \over \ln (1+\varepsilon_0)}\right)^2 \simeq r_0 \left({\varepsilon \over \varepsilon_0}\right)^2\;.
\end{equation}
As $r_0 \sim 10^{-104} \lambda_{\mathrm{Mpc}}^{-4}$ ($p=q=0$), a factor $|\varepsilon / \varepsilon_0|$ of the order $10^{52}$ could account for the observed magnetic fields even though under quite special conditions.


In what follows we shall look for solutions of the full Eq. (\ref{eq:evolF}), analyzing the effects of both the photon mass and the variation of the electromagnetic coupling. As before, we assume that the magnetic flux varies as a power of the conformal time -- a constant flux case is ruled out:
\begin{equation}
|\vec{F}_k| \propto \eta^m, \qquad m\neq 0\;,
\end{equation}
so that Eq. (\ref{eq:evolF}) turns into a first order equation for $\phi$, whose solution is
\begin{equation}
\label{eq:phisol1}
\phi(\eta)={n \over (1-m)m}+C \eta^{1-m}\;.
\end{equation}
The two terms of the solution are related to two models: the first corresponds to solution Eq. (\ref{eq:solF}) of the photon mass model, for which $\phi=1$ and $m$ is replaced by $m_\pm$; the second corresponds to the solutions Eqs. (\ref{eq:phisoldS}) and (\ref{eq:phisolRH}), with $m=p$ and $m=q$, respectively. Then, the second term corresponds to the variation of $\phi$ according to Eq. (\ref{eq:varphi}). The first term arises as the photon mass is constant for each period since $n$ and $m_\pm$ are constants.

In order to study the relevance of the first term, we may go back to Eq. (\ref{eq:evolF}) and assume that $\phi=\bar{\phi}$, where $\bar{\phi}$ is a constant value. The solution is given by Eq. (\ref{eq:solF}), apart from the changes
\begin{equation}
\label{eq:mphi}
m_\pm={1 \over 2}\left(1\pm \sqrt{1-4 n_{\bar{\phi}}}\right)\;
\end{equation}
with $n_{\bar{\phi}} \equiv n / \bar{\phi}$, so that
\begin{eqnarray}
\label{eq:pphi}
p={1 \over 2}\left(1- \sqrt{1-4 n_{\bar{\phi}_{dS}}}\right)\;, \\
\label{eq:qphi}
q={1 \over 2}\left(1+ \sqrt{1-4 n_{\bar{\phi}_{RH}}}\right)\;.
\end{eqnarray}
It is now clear that the influence of the mass term is related with the coupling. If $\bar{\phi} <1$, we obtain a larger effective $-n_{\bar{\phi}}$. This `step' variation of the coupling could considerably affect the photon mass model, as a result of the strong dependence of the ratio $r$ on the parameters $p$ e $q$. For numerical examples, see Table \ref{tab:rn0phi}. Magnetic field generation requires $\bar{\phi}<1$, meaning that the electromagnetic coupling was stronger than today's value. We note that this conclusion is the opposite from the analysis resulting from Eq. (\ref{eq:varphi}), which required considerable growth of the coupling.


We consider now the case in which $\phi$ varies linearly with the conformal time and assume that this behavior remains valid also for the RH phase. For each period, we have
\begin{equation}
\label{eq:philinear}
\phi(\eta)=\bar{\phi}+C_1 \eta\;,
\end{equation}
where $\bar{\phi}$ and $C$ are constants. The corresponding solutions of Eq. (\ref{eq:evolF}) are
\begin{equation}
\label{eq:solFhiper}
|\vec{F}_k|\propto \eta^{m_\pm}\,{\mathcal F_1}\left(m_\pm, {C_1 \eta \over \bar{\phi}}\right)\;,
\end{equation}
where $m_\pm=\frac{1}{2}(1\pm \sqrt{1-4n_{\bar{\phi}}})$, as before $n_{\bar{\phi}}\equiv n/\bar{\phi}$, and the function ${\mathcal F}$ denotes the hypergeometric function $_2F_1$ \cite{AS}:
\begin{equation}
\label{eq:defF}
{\mathcal F_1}\left(m, z\right) \equiv {}_2F_1\left(m,m,2 m, -z\right)\;.
\end{equation}
For $C=0$, we recover the case above as ${\mathcal F_1}\left(m,0\right)=1$. It is also possible to prove that, for $n=0$, the solution is, as expected, the same as Eq. (\ref{eq:solFlnphi}).

\begin{figure}[ht]
\centering
\leavevmode \epsfysize=7cm \epsfbox{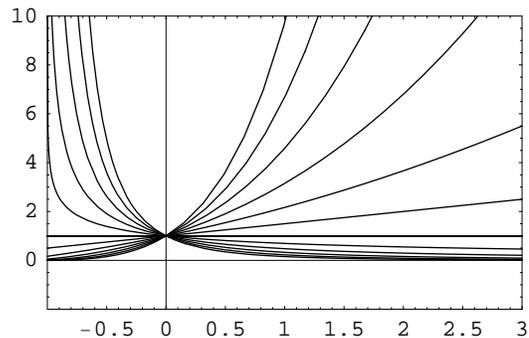}\\
\vskip 0.5cm
\caption{Graphic of ${\mathcal F_1}$ as a function of $C_1 \eta/ \bar{\phi}$ for integer values of $m$ $\{-3,\ldots,5\}$. The slope of the curves grows with $-m$: the extreme ones are $m=-3$ (positive slope) and $m=5$ (negative slope) and the limit curve for $m\to 0$ equals 1. The divergence on the left occurs for $\phi=0$.}
\label{fig:hiperm}
\end{figure}

The previous choice of the powers, which allows for $p<0$ and $q>1$, was made on the basis that it provides a noticeable growth of $|\vec{F}_k|$ in the dS and RH periods. We can see how that choice influences the factors ${\mathcal F_1}$ by looking at Figure~\ref{fig:hiperm}.

In the dS phase, the growth $a\propto -\frac{1}{\eta}$ cancels out the term $C_1 \eta$ and the value of $\phi$ by the end of the phase will be close to $\bar{\phi}$ (we are assuming $\bar{\phi}$ is not negligible). The interesting branch (see Figure~\ref{fig:hiperm}), which allows for an amplification when $C_1 \eta$ approaches zero for negative $m$, is the lower left one: function ${\mathcal F_1}$ starts from a small value and approaches 1, at the same time as $\phi$ grows from $\phi_0$ to $\bar{\phi}$.

In the RH phase, $a\propto \eta^2$, and $\bar{\phi}$ corresponds to the initial value of $\phi$. As that is also the dS period final value, we have $\bar{\phi}_{RH}\approx \bar{\phi}_{dS}\equiv \bar{\phi}$. The growing solution is now the one that moves away from vanishing $\eta$ along the branch that diverges at the left and it corresponds to a reduction of $\phi$. Assuming that only small variations of the electromagnetic coupling are admissible in the following evolution of the Universe, the final value of $\phi$ in the RH period must naturally be quite close to 1.

We can factorize the dependence of the solutions Eq. (\ref{eq:solFhiper}) with the conformal time and the variation of $\phi$. The expression for the ratio, now denoted by $r_\phi$, becomes
\begin{equation}
\label{eq:rhiper}
r_\phi \simeq r \left[{{\mathcal F_1}\left(q, {1 \over \bar{\phi}}-1\right) \over {\mathcal F_1}\left(p, {\phi_0 \over \bar{\phi}}-1\right)}\right]^2\;,
\end{equation}
where $r$ is given by Eq. (\ref{eq:r}) with the values depicted in Eqs. (\ref{eq:pphi}) and (\ref{eq:qphi}).

Let us now estimate $r_\phi$. We assume that $p=-2$ and $q=1$ (different values for $q$ lead to similar results). 
Interesting variations of $\phi$ include growth at the dS period and attenuation at the RH period. With maximum 
at $\bar{\phi}$, the approximations $\bar{\phi}\gg \phi_0$ and $\bar{\phi}\gg 1$ imply that
\begin{equation}
\label{eq:f1part}
\left[{{\mathcal F_1}\left(1, {1 \over \bar{\phi}}-1\right) \over {\mathcal F_1}\left(-2, {\phi_0 \over \bar{\phi}}-1\right)}\right]^2 
\simeq {\log^2{{1 \over \bar{\phi}}} \over {1 \over 36}}= 36 \log^2{\bar{\phi}}\;,
\end{equation}
meaning that, even with large variations of $\phi$, the resulting amplification is very reduced.


Actually, the analysis of the function $\mathcal F_1$ deserves more attention. The transition between 
the curves for the various negative values of $m$ is not smooth; for semi-integer values of $m$, the 
function diverges. These cases might be interesting if they correspond to considerable amplification 
of the fluctuations. Their evolution implies for the photon mass 
\begin{equation}
n_{\bar{\phi}_{dS}}=-[(j+1)^2-\frac{1}{4}]\;,
\end{equation}
as $p=-\left(j+\frac{1}{2}\right)$ with $j=0,1,2,\ldots$; however, according to Ref. \cite{BM},
\begin{equation}
n=-{M_{dS}^2 \over H_{dS}^2}
\end{equation}
for the dS phase, $M_{dS}$ being the mass in the Lagrangian and $H_{dS}$ the expansion rate during inflation, so that we have
\begin{equation}
M_{dS}=\sqrt{\left((j+1)^2-\frac{1}{4}\right) \bar{\phi}_{dS}} \times H_{dS}\;.
\end{equation} 
As the evolution of $\mathcal F_1$ is opposite for values of $p$ around the divergence point, 
${\mathcal F_1}^2$ has a root for every divergence. For values of $p$ and initial $C_1 \eta/ \bar{\phi}$ 
near a root of ${\mathcal F_1}^2$ an amplification may arise. Of course, this is possible only if those 
values are conveniently fine-tuned. Due to the linearity of ${\mathcal F_1}$ around the roots, the precision 
of the tuning will control the amount of amplification. If $C_1 \eta/ \bar{\phi}$ and $p$ are both tuned 
to the root with a precision of $k$ digits, the factor ${\mathcal F_1}^2$ will cause an amplification of the order $10^{2k}$.

We consider now the quadratic case
\begin{equation}
\phi(\eta)=\bar{\phi}+C_2 \eta^2 \;,
\end{equation}
allowed by Eq. (\ref{eq:phic1c2}), which brings nothing new when compared to the linear case of Eq. (\ref{eq:philinear}). 
The solutions of Eq. (\ref{eq:evolF}) are now
\begin{equation}
|\vec{F}_k|\propto \eta^{m_\pm}\,{\mathcal F_2}\left(m_\pm, {C_2 \eta^2 \over \bar{\phi}}\right)\;,
\end{equation}
where the function ${\mathcal F_2}$ denotes \cite{AS}:
\begin{equation}
{\mathcal F_2}\left(m, {z}\right) \equiv {}_2F_1\left({m \over 2},{1+m \over 2},{1 \over 2}+m, -z\right)\;.
\end{equation}

The behavior of ${\mathcal F_2}$ is analogous to the one of ${\mathcal F_1}$, given in Figure~\ref{fig:hiperm}, but the values of the former are closer to unity. The equivalent to Eq. (\ref{eq:f1part}) is now
\begin{equation}
\left[{{\mathcal F_2}\left(1, {1 \over \bar{\phi}}-1\right) \over {\mathcal F_2}\left(-2, {\phi_0 \over \bar{\phi}}-1\right)}\right]^2 \simeq {{1 \over 4}\log^2{{1 \over 4\bar{\phi}}} \over {4 \over 9}}= {9 \over 16} \log^2{4 \bar{\phi}}\;,
\end{equation}
with similar conclusions.

Moreover, in what refers to the divergence points of ${\mathcal F_2}$, the conclusions are equivalent to the ones concerning ${\mathcal F_1}$.

Notice that if we have had considered the case of Eq. (\ref{eq:phic3}), corresponding to $V(\phi)=0$, the 
conclusions would be also analogous.

\section{Conclusions}

The origin of large-scale magnetic fields remains still an intriguing issue in astrophysics. The galactic 
dynamo process is a feasible mechanism for the amplification fields over galaxy scales, provided seed magnetic 
fields are generated before the Universe becomes a good conductor. It is evident that inflation is a quite 
interesting process of stretching quantum fluctuations of the electromagnetic field till galactic scales 
provided the conformal symmetry of electromagnetism is broken.

In this work, we have considered the question of generation of seed magnetic fields in the context of a 
model where electromagnetism is coupled to a scalar field. We have shown that the isolated influence 
of electromagnetic coupling variations cannot produce the required seed magnetic fields. The growth 
factor needed for the coupling, assuming that the magnetic flux grows as a power of the scale factor, 
is in the range $10^{27}$ to $10^{38}$ in order to allow for $r \sim 10^{-34}$.

We have shown that, when considering the photon mass model, a constant value of the electromagnetic 
coupling greater than the present value ($\bar{\phi} <1$) would result in a relevant strengthening of the amplification process.

On the other hand, in the cases we have analyzed in detail, namely $\phi \sim \eta$ and $\phi \sim \eta^2$, 
an additional but hardly relevant amplification $\sim \log^2 \bar{\phi}$ arises. We stress that the solutions 
for the variation of the electromagnetic coupling may naturally lead to $\bar{\phi}<1$, where $\bar{\phi}$ is 
the inverse of the square of the electromagnetic coupling at the end of the de Sitter period.

\vfill


\end{document}